\documentclass[twocolumn,prl]{revtex4}
\usepackage{amsfonts}
\usepackage{amsmath,amsbsy,amssymb,graphicx}

\def\beginABC{\begin{subequations}}
\def\endABC{\end{subequations}}

\begin{document}

\title{{\Large Quasi-Phase Transition and Many-Spin Kondo Effects\ in
Graphene Nanodisk}}
\author{Motohiko Ezawa}
\affiliation{Department of Applied Physics, University of Tokyo, Hongo 7-3-1, 113-8656,
Japan }

\begin{abstract}
The trigonal zigzag nanodisk with size $N$ has $N$ localized spins. We
investigate its thermodynamical properties with and without external leads.
Leads are made of zigzag graphene nanoribbons or ordinary metallic wires.
There exists a quasi-phase transition between the quasi-ferromagnet and
quasi-paramagnet states, as signaled by a sharp peak in the specific heat
and in the susceptability. Lead effects are described by the many-spin Kondo
Hamiltonian. A new peak emerges in the specific heat. Furthermore, the band
width of free electrons in metallic leads becomes narrower. By investigating
the spin-spin correlation it is argued that free electrons in the lead form
spin-singlets with electrons in the nanodisk. They are indications of
many-spin Kondo effects.
\end{abstract}

\maketitle


\address{{\normalsize Department of Applied Physics, University of Tokyo, Hongo
7-3-1, 113-8656, Japan }}

\textit{Introduction:} Graphene nanostructure\cite{GraphEx} has the
potential for future application in nanoelectronics and spintronics. In
particular, much attention is now focused on graphene nanoribbons\cite%
{Nanoribbon} due to almost flat low-energy band at the Fermi level depending
on the edge states. Another basic element of graphene derivatives is a
graphene nanodisk\cite%
{EzawaDisk,EzawaPhysica,Fernandez,Hod,Wang,EzawaCoulomb}. It is a
nanometer-scale disk-like material which has a closed edge. There are many
type of nanodisks, among which the trigonal zigzag nanodisk is prominent in
its electronic and magnetic properties because there exist $N$-fold
degenerated half-filled zero-energy states when its size is $N$ (Fig.\ref%
{FigNanodisk}).

In this paper we explore thermodynamical properties of the trigonal zigzag
nanodisk. The system is well approximated by the infinite-range Heisenberg
model. It is exactly solvable. A sharp peak emerges in the specific heat and
in the susceptibility, which we interpret as a quasi-phase transition
between the quasi-ferromagnet and quasi-paramagnet states. We then
investigate a nanodisk-lead system, where the lead is made of a zigzag
graphene nanoribbon or an ordinary metallic wire. We refer to it as a
graphene lead or a metallic lead.

It is shown that lead effects are described by the many-spin Kondo
Hamiltonian. Electron spins in the nanodisk and the lead orient into the
opposite directions to lower the coupling energy. A new peak appears around
a certain temperature ($T=T_{\text{K}}$) in the specific heat but not in the
susceptibility for small size nanodisks. The internal energy decreases near
the peak, and the band width of free electrons in the lead becomes narrower
in the instance of the metallic lead. Furthermore, the spin-spin correlation
takes the maximum value at $T=0$, remains almost constant for $T\lesssim T_{%
\text{K}}$, and then decreases monotonically as $T$ increases. We interpret
these phenomena to mean that free electrons in the lead are consumed to make
spin-singlets with electrons in the nanodisk. It is intriguing that all
electrons (only a few portion of electrons) in the nanodisk are engaged in
the case of the graphene (metallic) lead. They are indications of Kondo
effects due to the Kondo coupling between electrons in the lead and the
nanodisk.

\begin{figure}[t]
\includegraphics[width=0.44\textwidth]{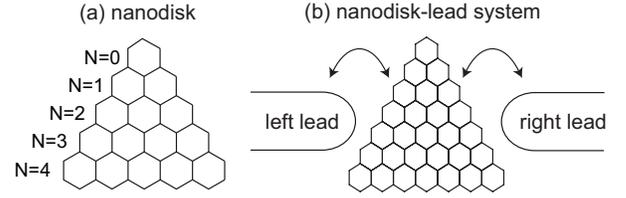}
\caption{(a) Trigonal zigzag nanodisks. The size parameter $N$ is defined in
this way. The number of carbon atoms is given by $N_{\text{C}}=N^{2}+6N+6$.
(b) The nanodisk-lead system. The nanodisk with $N=7$ is connected to the
right and left leads.}
\label{FigNanodisk}
\end{figure}

\textit{Quasi-Ferromagnet:} The size-$N$ zigzag trigonal nanodisk has $N$%
-fold degenerated zero-energy states\cite{EzawaDisk}, where the gap energy
is as large as a few eV. Hence it is a good approximation to investigate the
electron-electron interaction physics only in the zero-energy sector, by
projecting the system to the subspace made of those zero-energy states. The
zero-energy sector consists of $N$ orthonormal states $|f_{\alpha }\rangle $%
, $\alpha =1,2,\cdots ,N$, together with the SU(N) symmetry. Let $U_{\alpha
\beta }$ and $J_{\alpha \beta }$ be the Coulomb energy and the exchange
energy between electrons in the states $|f_{\alpha }\rangle $ and $|f_{\beta
}\rangle $. It follows\cite{EzawaCoulomb} that $J_{\alpha \beta }\approx
U_{\alpha \beta }$ and that all $J_{\alpha \beta }$ are of the same order of
magnitude for any pair of $\alpha $ and $\beta $, implying that the SU(N)
symmetry is broken but not so strongly. It is a good approximation to start
with the exact SU($N$) symmetry, by replacing $U_{\alpha \beta }$ and $%
J_{\alpha \beta }$ with their averages $U$ and $J$, respectively. Then, the
zero-energy sector is described by the Hamiltonian\cite{EzawaSpin}, 
\begin{equation}
H_{\text{D}}=-J\mathbf{S}_{\text{tot}}^{2}+\frac{1}{2}U^{\prime }n_{\text{tot%
}}^{2}+\left( \frac{U}{2}+J\right) n_{\text{tot}},  \label{HamilDx}
\end{equation}%
with $U^{\prime }\equiv U-\frac{1}{2}J$ and $J\approx U$, where $\mathbf{S}_{%
\text{tot}}=\sum_{\alpha }\mathbf{S}\left( \alpha \right) $ is the total
spin, and $n_{\text{tot}}=\sum_{\alpha }n\left( \alpha \right) $ is the
total electron number. Here, $n\left( \alpha \right) =d_{\sigma }^{\dag
}(\alpha )d_{\sigma }(\alpha )$, and $\mathbf{S}(\alpha )=\frac{1}{2}%
d_{\sigma }^{\dag }(\alpha )\mathbf{\tau }_{\sigma \sigma ^{\prime
}}d_{\sigma ^{\prime }}(\alpha )$ with $d_{\sigma }(\alpha )$ the
annihilation operator of electron with spin $\sigma =\uparrow ,\downarrow $
in the state $|f_{\alpha }\rangle $: $\mathbf{\tau }$ is the Pauli matrix.
We call $\mathbf{S}_{\text{tot}}$ the nanodisk spin.

Apart from an irrelevant constant, the Hamiltonian (\ref{HamilDx}) is the
infinite-range Heisenberg model, $H_{\text{S}}=-J\mathbf{S}_{\text{tot}}^{2}$%
, in the half-filled sector with $n_{\text{tot}}=N$. The nanodisk spin takes
the maximum value $S_{g}=\sqrt{N/2\left( N/2+1\right) }$ in the ground
state, where all spins are spontaneously polarized. The nanodisk spin system
exhibits a strong ferromagnetic order due to a large exchange interaction.
The relaxation time is finite but quite large even if the size $N$ is small.
We have called such a system quasi-ferromagnet\cite{EzawaDisk}.

\textit{Thermodynamical Properties:} The infinite-range Heisenberg model is
exactly diagonalizable, $H_{\text{S}}|\Psi _{S}\rangle =E_{S}|\Psi
_{S}\rangle $, with $E_{S}=-JS(S+1)$, where $S$ takes half-integer or
integer values from $N/2$ down to $1/2$ or $0$, depending on whether $N$ is
odd or even. The total degeneracy of the energy level $E_{S}$ is $\left(
2S+1\right) g_{N}(S)$ with $g_{N}\left( N/2-q\right)
=_{N}\!\!C_{q}-_{N}\!\!C_{q-1}$.

\begin{figure}[t]
\includegraphics[width=0.5\textwidth]{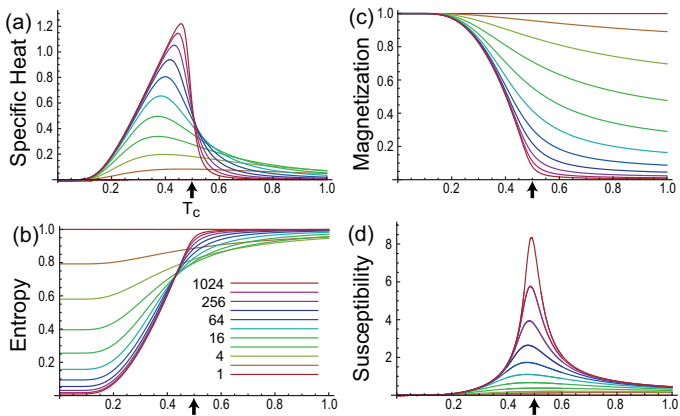}
\caption{Thermodynamical properties of the nanodisk without leads. (a) The
specific heat $C$ in unit of $k_{\text{B}}N$. (b) The entropy $S$ in unit of 
$k_{\text{B}}N\log 2$. (c) The magnetization $\left\langle \mathbf{S}_{\text{%
tot}}^{2}\right\rangle $ in unit of $S_{g}^{2}$. (d) The susceptibility $%
\protect\chi $ in unit of $S_{g}$. The size is $N=1,2,2^{2},\cdots 2^{10}$.
The horizontal axis stands for the temperature $T$ in unit of $JN/k_{\text{B}%
}$. The arrow represents the phase transition point $T_{c}$ in the limit $%
N\rightarrow \infty $.}
\label{FigThermo}
\end{figure}

We have a complete set of the eigenenergies together with their
degeneracies. The partition function of the nanodisk with size $N$ is
exactly calculable,%
\begin{equation}
Z_{\text{S}}=\sum_{S}\left( 2S+1\right) g_{N}(S)e^{-\beta JS(S+1)}.
\end{equation}%
According to the standard procedure we can evaluate the specific heat $C_{%
\text{N}}(T)$, the entropy $S_{\text{N}}(T)$, the magnetization $%
\left\langle \mathbf{S}_{\text{tot}}^{2}\right\rangle $ and the
susceptibility $\chi =\frac{1}{k_{\text{B}}T}\big(\left\langle \mathbf{S}_{%
\text{tot}}^{2}\right\rangle -\left\langle \mathbf{S}_{\text{tot}%
}\right\rangle ^{2}\big)$ from this partition function. The entropy is given
by $S_{\text{N}}\left( 0\right) =k_{\text{B}}\log (N+1)$ at zero
temperature, corresponding to the ground state multiplicity $N+1$. We
display them in Fig.\ref{FigThermo} for size $N=1,2,2^{2},\cdots 2^{10}$.

There appear singularities in thermodynamical quantities as $N\rightarrow
\infty $, which represent a phase transition at $T_{c}\equiv JN/2k_{\text{B}%
} $ between the ferromagnet and paramagnet states (Fig.\ref{FigThermo}). For
finite $N$, there are steep changes around $T_{c}$, though they are not
singularities. It is not a phase transition. However, it would be reasonable
to call it a quasi-phase transition between the quasi-ferromagnet and
quasi-paramagnet states. Such a quasi-phase transition is manifest even in
finite systems with $N=100$ $\sim $ $1000$.

The specific heat and the magnetization take nonzero-values for $T>T_{c}$
[Fig.\ref{FigThermo}(a),(c)], which is zero in the limit $N\rightarrow
\infty $. The entropy for $T>T_{c}$ is lower than that of the paramagnet
[Fig.\ref{FigThermo}(b)]. These results mean the existence of some
correlations in the quasi-paramagnet state. The maximum value of the
susceptibility increases linearly as $N$ becomes large. It is an indicator
of the quasi-phase transition.

\textit{Many-Spin Kondo Hamiltonian:} We proceed to investigate how
thermodynamical properties of the nanodisk is affected by the attachment of
the leads. Though there are two leads attached to a nanodisk, the lead
Hamiltonian $H_{\text{L}}$ and the transfer Hamiltonian $H_{\text{T}}$ are
expressed as if there were a single lead after a certain transformation\cite%
{EzawaCoulomb},\beginABC\label{HamilTLR}%
\begin{align}
H_{\text{L}}& =\sum_{k\sigma }\varepsilon \left( k\right) c_{k\sigma
}^{\dagger }c_{k\sigma },  \label{HamilL} \\
H_{\text{T}}& =\tilde{t}\sum_{k\sigma }\sum_{\alpha }\left( c_{k\sigma
}^{\dagger }d_{\sigma }(\alpha )+d_{\sigma }^{\dagger }(\alpha )c_{k\sigma
}\right) ,  \label{HamilT}
\end{align}%
\endABC where $c_{k\sigma }$ is the annihilation operator of electron in the
lead with the wave number $k$ and the dispersion relation $\varepsilon
\left( k\right) $.

When charges transfer between the nanodisk and the leads, the total electron
number $n_{\text{tot}}$ is no longer fixed in the nanodisk. However, the
nanodisk remains to be half-filled, when a charge transfers from the lead to
the nanodisk and then transfers back from the nanodisk to the lead. The
process is the second order effect in the tunneling coupling constant $%
\tilde{t}$. We derive the effective Hamiltonian for such a process.

The total Hamiltonian is $H=H_{\text{D}}+H_{\text{L}}+H_{\text{T}}$. We take 
$H_{0}=H_{\text{D}}+H_{\text{L}}$ as the unperturbed term\ and $H_{\text{T}}$
as the perturbation term. Note that $U\gg \tilde{t}$. We make a canonical
transformation known as the Schrieffer-Wolff transformation\cite%
{SchriefferWolff} to eliminate $H_{\text{T}}$, $H\rightarrow \widetilde{H}%
=e^{iG}He^{-iG}$, with $G$ the generator satisfying $H_{\text{T}}+\frac{i}{2}%
\left[ G,H_{0}\right] =0$. The dominant contribution comes from the Fermi
surface, $\varepsilon \left( k\right) =\varepsilon _{\text{F}}$. We assume
the symmetric condition $\varepsilon _{\text{F}}=U+\frac{1}{2}U^{\prime }$
with respect to the Fermi energy. Then, after a straightforward calculation,
we obtain $\widetilde{H}=H_{\text{D}}+H_{\text{L}}+H_{\text{K}}+O(\tilde{t}%
^{3})$, where $H_{\text{K}}$ is the second order term in $\tilde{t}$. It is
the many-spin Kondo Hamiltonian,%
\begin{equation}
H_{\text{K}}=J_{\text{K}}\sum_{kk^{\prime }\sigma \sigma ^{\prime
}}c_{k\sigma }^{\dagger }\mathbf{\tau }_{\sigma \sigma ^{\prime
}}c_{k^{\prime }\sigma ^{\prime }}\cdot \mathbf{S}_{\text{tot}},
\label{Kondo}
\end{equation}%
with the Kondo coupling constant $J_{\text{K}}=8\tilde{t}^{2}/U^{\prime }$.
The difference between the above many-spin Kondo Hamiltonian and the
ordinary Kondo Hamiltonian is whether the local spin is given by the
summation over many spins $\mathbf{S}_{\text{tot}}$ or a single spin $%
\mathbf{S}$. Note that $\mathbf{S}_{\text{tot}}^{2}$ is a dynamical variable
but $\mathbf{S}^{2}$ is not, $\mathbf{S}^{2}=3/4$.

\textit{Functional Integration:} The total Hamiltonian is now given by $H_{%
\text{eff}}=H_{\text{S}}+H_{\text{L}}+H_{\text{K}}$ at half filling. We
define the spinor $\psi =(c_{\uparrow },c_{\downarrow })^{t}$. The partition
function in the Matsubara form is given in terms of the Hamiltonian density $%
\mathcal{H}_{\text{eff}}$ as%
\begin{align}
Z &=\text{Tr}_{S}\int \mathcal{D}\psi \mathcal{D}\psi ^{\dagger }\exp \left[
-\int_{0}^{\beta }d\tau \int dx\,\left( \psi ^{\dagger }\partial _{\tau
}\psi +\mathcal{H}_{\text{eff}}\right) \right]  \notag \\
&=\text{Tr}_{S}\left[ \exp \left( -\beta H_{\text{S}}\right) Z_{\text{K}}%
\right] ,  \label{TotalZ}
\end{align}%
with $Z_{\text{K}}=\int \mathcal{D}\psi \mathcal{D}\psi ^{\dagger }\exp %
\left[ -S_{\text{K}}\right] $, where $S_{\text{K}}$ is the action $S_{\text{K%
}}=\int_{0}^{\beta }d\tau \int dx\,\left( \psi ^{\dagger }\partial _{\tau
}\psi +\mathcal{H}_{\text{L}}+\mathcal{H}_{\text{K}}\right) $. We first
perform a functional integral over the lead electron's degree of freedom in $%
Z_{\text{K}}$, and then summed up the nanodisk spin in (\ref{TotalZ}).

Because an electron in the lead is constrained within a very narrow region,
it is a good approximation to neglect momentum scatterings,%
\begin{equation}
H_{\text{K}}\simeq 2J_{\text{K}}\mathbf{s}\cdot \mathbf{S}_{\text{tot}},
\label{KondoCoupl}
\end{equation}%
where $\mathbf{s}=\frac{1}{2}\sum_{k\sigma \sigma ^{\prime }}c_{k\sigma
}^{\dagger }\mathbf{\tau }_{\sigma \sigma ^{\prime }}c_{k\sigma ^{\prime }}$
is the electron spin in the lead. The action $S_{\text{K}}$ is summarized as 
\begin{equation}
S_{\text{K}}=\int \frac{d\omega }{2\pi }\sum_{k}\psi ^{\dagger }\left(
k\right) M\left( k\right) \psi \left( k\right) ,
\end{equation}%
with $M\left( k\right) =-[i\omega -\varepsilon \left( k\right) ]+J_{\text{K}}%
\mathbf{\tau }\cdot \mathbf{S}_{\text{tot}}$. Performing the functional
integration we find $Z_{\text{K}}=$Det$[M]=\exp \left[ -\beta F_{\text{K}}%
\right] $, where $F_{\text{K}}$\ is the Helmholtz free energy, 
\begin{equation}
F_{\text{K}}=-\frac{1}{2\beta }\sum_{k}\ln \left[ \cosh (\beta J_{\text{K}%
}\left\vert \mathbf{S}_{\text{tot}}\right\vert )+\cosh (\beta \varepsilon
\left( k\right) )\right] .  \label{KFree}
\end{equation}%
This is reduced to the well-known formula for free electrons with the
dispersion relation $\varepsilon (k)$ for $J_{\text{K}}=0$.

\begin{figure}[t]
\includegraphics[width=0.5\textwidth]{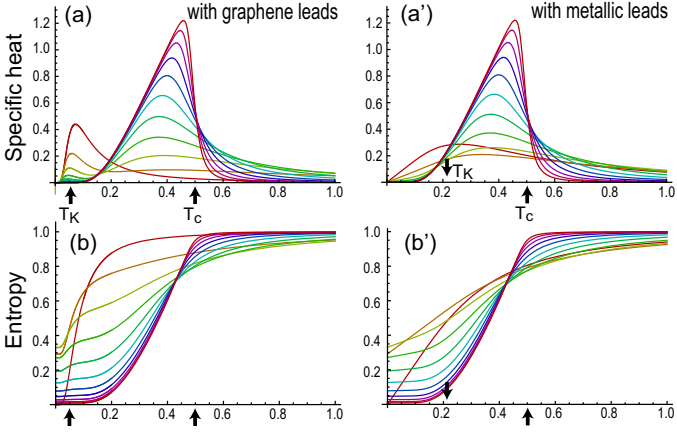}
\caption{Thermodynamical properties of the nanodisk (a,b) with graphene
leads and (a',b') with metallic leads. See the caption of Fig.\protect\ref%
{FigThermo}. We have set $J_{\text{K}}/J=0.2$ and $D=2k_{\text{B}}T_{c}$. A
new peak appears around $T_{\text{K}}$ in the specific heat.}
\label{FigLthermo}
\end{figure}

\textit{Zigzag Graphene Nanoribbon Leads:} We first consider the system
where the leads are made of zigzag graphene nanoribbons. Owing to the flat
band at the zero energy, $\varepsilon \left( k\right) =0$, the result of the
functional integration (\ref{KFree}) is quite simple,%
\begin{equation}
F_{\text{K}}=-\frac{1}{\beta }\ln \cosh [\beta J_{\text{K}}\left\vert S_{%
\text{tot}}\right\vert /2].
\end{equation}%
The effective Hamiltonian for the nanodisk spin is $H_{\text{S}}+F_{\text{K}}
$. The lead effect is to make the effective spin stiffness larger and the
ferromagnet more rigid.

The trace over the nanodisk spin is carried out in (\ref{TotalZ}),%
\begin{equation}
Z=\sum_{S}\left( 2S+1\right) g_{N}(S)e^{-\beta JS(S+1)}\cosh \Big(\frac{%
\beta J_{\text{K}}}{2}\sqrt{S(S+1)}\Big).
\end{equation}%
We compare thermodynamical properties of the nanodisk with leads (Fig.\ref%
{FigLthermo}) and without leads (Fig.\ref{FigThermo}). The magnetization $%
\left\langle \mathbf{S}_{\text{tot}}^{2}\right\rangle $ and the
susceptibility $\chi $ are found to be indistinguishable from those of the
nanodisk without leads (Fig.\ref{FigThermo}). In Fig.\ref{FigLthermo}, we
show the specific heat $C_{\text{G}}(T)$ and the entropy $S_{\text{G}}(T)$
for various size $N$. The significant feature is the appearance of a new
peak in the specific heat at $T=T_{\text{K}}\approx (J_{\text{K}}/2J)T_{c}$,
though it disappears for large $N$. We examine the internal energy $E_{\text{%
G}}(T)$, which is found to decrease around $T_{\text{K}}$ (Fig.\ref%
{FigEnergy}). Near zero temperature it reads%
\begin{equation}
E_{\text{G}}(T)\simeq -JS_{g}^{2}-\frac{J_{\text{K}}}{2}S_{g}+J_{\text{K}%
}S_{g}e^{-\beta J_{\text{K}}S_{g}}.  \label{EnergRibbo}
\end{equation}%
The first term ($\propto J$) represents the energy stabilization due to the
ferromagnetic order present in the nanodisk without leads, while the second
term ($\propto J_{\text{K}}$) represents the one due to the Kondo coupling.
Furthermore, it follows that the entropy is reduced at zero temperature as $%
S_{\text{G}}\left( 0\right) -S_{\text{N}}\left( 0\right) =-k_{\text{B}}\log
2 $, as implies that the ground state multiplicity at the zero temperature
is just one half of that of the system without leads.

\begin{figure}[tb]
\includegraphics[width=0.5\textwidth]{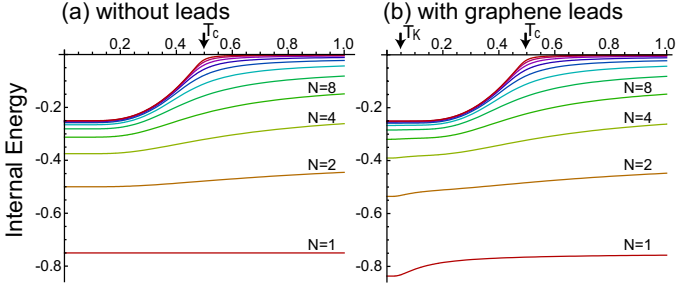}
\caption{{}The internal energy $E$ in unit of $Nk_{\text{B}}$ \ for the
nanodisk (a) without leads and (b) with graphene leads. See the caption of
Fig.\protect\ref{FigThermo}. (a) The energy descreases except for $N=1$ as
the temperature decreases, which represents the ferromagnetic order. (b)
There exists an additional energy decrease around $T_{\text{K}}$, which is
prominent for $N=1$, attributed to the Kondo effect.}
\label{FigEnergy}
\end{figure}

\begin{figure}[t]
\includegraphics[width=0.5\textwidth]{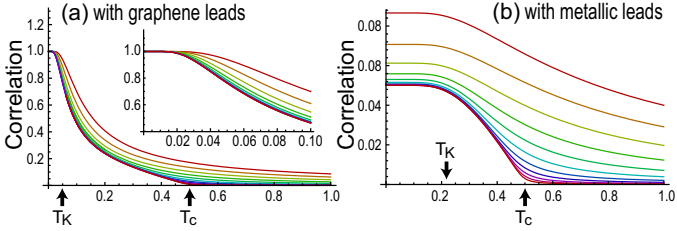}
\caption{{}The spin-spin correlation in unit of $S_{g}/2$. See the caption
of Fig.\protect\ref{FigLthermo}. The correlation occurs due to the Kondo
coupling. It is almost constant for $T\lesssim T_{\text{K}}$, and decreases
as $T$ increases.}
\label{FigsS}
\end{figure}

The spin-spin correlation $|\langle \mathbf{s}\cdot \mathbf{S}_{\text{tot}%
}\rangle |$ is calculated based on the partition function (\ref{TotalZ}) and
shown in Fig.\ref{FigsS}(a). Near zero temperature we find%
\begin{equation}
|\langle \mathbf{s}\cdot \mathbf{S}_{\text{tot}}\rangle |\simeq \frac{1}{2}%
S_{g}\tanh \left[ \beta J_{\text{K}}S_{g}/2\right] .
\end{equation}%
It takes the maximum value $S_{g}/2$ at $T=0$, and remains almost constant
for $T\lesssim T_{\text{K}}$, and then monotonically decreases as $T$
increases. Finally, it almost vanishes in the quasi-paramagnet phase for
large $N$ since $\langle \mathbf{S}_{\text{tot}}\rangle \approx 0$. We may
interpret these phenomena as follows. Electrons in the lead and the nanodisk
form spin-singlet states to lower the coupling energy (\ref{KondoCoupl}).
The singlet state is rather tight for $T\lesssim T_{\text{K}}$, but
thermally broken as $T$ increases.

All these features indicate the occurrence of the Kondo effect due to the
coupling between the spins in the nanodisk and the leads.

\textit{Metallic Leads:} Next we consider the system comprised of metallic
leads with a constant energy density, $\rho \left( \varepsilon \right) =\rho 
$ for $\left\vert \varepsilon \right\vert <D$ and $\rho \left( \varepsilon
\right) =0$ for $\left\vert \varepsilon \right\vert >D$. We change the
momentum integration into the energy integration in (\ref{KFree}),%
\begin{equation}
F_{\text{K}}=-\frac{\rho }{2\beta }\int_{-D}^{D}d\varepsilon \ln \big[\cosh
\beta J_{\text{K}}\left\vert \mathbf{S}_{\text{tot}}\right\vert +\cosh \beta
\varepsilon \big].
\end{equation}%
The free energy is given by $F_{\text{K}}=-\rho \beta ^{-2}\big[\beta
^{2}DJ_{\text{K}}\left\vert \mathbf{S}_{\text{tot}}\right\vert -2\beta D\log
2+$Li$_{2}\big(-e^{\beta \left( D-J_{\text{K}}\left\vert \mathbf{S}_{\text{%
tot}}\right\vert \right) }\big)-$Li$_{2}\big(-e^{-\beta \left( D+J_{\text{K}%
}\left\vert \mathbf{S}_{\text{tot}}\right\vert \right) }\big)\big]$, where Li%
$_{2}\left[ x\right] $ is the dilogarithm function\cite{Abramowitz}. It is
reduced to the free energy of the nanodisk with graphene leads in the limit $%
D\rightarrow 0$ with $\rho =1/2D$.

The magnetization $\left\langle \mathbf{S}_{\text{tot}}^{2}\right\rangle $
and the susceptibility $\chi $ are found to be indistinguishable from those
in the case of no leads (Fig.\ref{FigThermo}). The relation $S_{\text{M}%
}\left( 0\right) -S_{\text{N}}\left( 0\right) =-k_{\text{B}}\log 2$ holds
for the entropy precisely as in the case of graphene leads. In Fig.\ref%
{FigLthermo}, we show the specific heat $C_{\text{M}}(T)$ and the entropy $%
S_{\text{M}}(T)$ for various size $N$. A broad peak appears at $T=T_{\text{K}%
}$ in the specific heat. Using the asymptotic behaviors\cite{Abramowitz} of
Li$_{2}\left[ x\right] $, we obtain the free energy in low temperature
regime as%
\begin{equation}
\!F_{\text{K}}\simeq \frac{-\rho }{\beta }\Big[\beta DJ_{\text{K}}\left\vert 
\mathbf{S}_{\text{tot}}\right\vert -2D\log 2+\frac{\pi ^{2}}{6\beta }+\frac{%
\beta }{2}\left( D-J_{\text{K}}\left\vert \mathbf{S}_{\text{tot}}\right\vert
\right) ^{2}\Big],
\end{equation}%
which is valid up to the terms in the order of $e^{-\beta D}$. The entropy $%
S_{\text{M}}\left( T\right) $, the specific heat $C_{\text{M}}\left(
T\right) $ and the internal energy $E_{\text{M}}\left( T\right) $ read as
follows,%
\begin{equation}
\!S_{\text{M}}\left( T\right) \simeq k_{\text{B}}\log \frac{N+1}{2}+\frac{%
\pi ^{2}}{3}\rho k_{\text{B}}^{2}T,\quad C_{\text{M}}\left( T\right) \simeq 
\frac{\pi ^{2}}{3}\rho k_{\text{B}}^{2}T,
\end{equation}%
and $E_{\text{M}}\left( T\right) =E_{\text{G}}\left( T\right) +\Delta E(T)$
with%
\begin{equation}
\Delta E(T)\simeq -\frac{\rho }{2}\left( D-J_{\text{K}}S_{g}\right) ^{2}+%
\frac{\pi ^{2}}{6}\rho \left( k_{\text{B}}T\right) ^{2}.
\end{equation}%
All terms proportional to $\rho $ have arisen from free electrons in the
metallic lead. The internal energy $E_{\text{M}}\left( T\right) $ consists
of two terms: $E_{\text{G}}\left( T\right) $ is identical to the energy (\ref%
{EnergRibbo}) for the nanodisk with graphene leads, and $\Delta E(T)$\ is
the energy of the metallic lead. The first term of $\Delta E(T)$ shows that
the band width of free electrons in the lead becomes narrower due to the
Kondo coupling. We may interpret that $n$ free electrons with 
\begin{equation}
n=\rho J_{\text{K}}S_{g}  \label{MNum}
\end{equation}%
are consumed to make spin-coupling with electrons in the nanodisk. The
second term is the thermal energy of free electrons in the metallic lead.

We show the spin-spin correlation $|\langle \mathbf{s}\cdot \mathbf{S}_{%
\text{tot}}\rangle |$ in Fig.\ref{FigsS}(b). The overall features are the
same as for the nanodisk with graphene leads. However, there are new
features. First of all, the value of correlation is quite small. This is
because spin-singlets are formed only by a small portion of electrons in the
metallic lead which are near the Fermi level. We expect that this number
density is given by (\ref{MNum}). Indeed, it is observed that $|\langle 
\mathbf{s}\cdot \mathbf{S}_{\text{tot}}\rangle |\rightarrow \frac{1}{2}\rho
J_{\text{K}}S_{g}$ at $T=0$ as $N\rightarrow \infty $ [Fig.\ref{FigsS}(b)].

In this paper we have investigated thermodynamical properties of a zigzag
graphene nanodisk without and with leads. The lead effects are summarized by
the many-spin Kondo Hamiltonian. One effect is to enhance the ferromagnetic
order. This result is important to manufacture spintronic circuits by
connecting leads in nanodevices\cite{EzawaSpin}. We have shown various
thermodynamical results indicating many-spin Kondo effects.

I am very much grateful to N. Nagaosa for many fruitful discussions on the
subject. This work was supported in part by Grants-in-Aid for Scientific
Research from the Ministry of Education, Science, Sports and Culture No.
20840011.


\begin{thebibliography}{99}
\bibitem{GraphEx} K. S. Novoselov, \textit{et al.}, Science \textbf{306},
666 (2004). K. S. Novoselov, \textit{et al.}, Nature \textbf{438}, 197
(2005). Y. Zhang, \textit{et al.}, Nature \textbf{438}, 201 (2005).

\bibitem{Nanoribbon} M. Fujita, \textit{et al.}, J. Phys. Soc. Jpn. \textbf{%
65}, 1920 (1996). M. Ezawa, Phys. Rev. B, \textbf{73}, 045432 (2006). L.
Brey, and H. A. Fertig, Phys. Rev. B, \textbf{73}, 235411 (2006). F. Mu\~{n}%
oz-Rojas, \textit{et al.}, Phys. Rev. B, \textbf{74}, 195417 (2006). Y. -W
Son, M. L. Cohen, and S. G. Louie, Phys. Rev. Lett., \textbf{97}, 216803
(2006). V. Barone, O. Hod, and G. E. Scuseria, Nano Lett., \textbf{6}, 2748
(2006). M. Y. Han, \textit{et al.}, Phys. Rev. Lett., \textbf{98}, 206805
(2007).

\bibitem{EzawaPhysica} M. Ezawa, Physica Status Solidi (c) \textbf{4}, No.2,
489 (2007).

\bibitem{EzawaDisk} M. Ezawa, Phys. Rev. B \textbf{76}, 245415 (2007): M.
Ezawa, Physica E \textbf{40}, 1421-1423 (2008).

\bibitem{Fernandez} J. Fern\'{a}ndez-Rossier, and J. J. Palacios, Phys. Rev.
Lett. \textbf{99}, 177204 (2007).

\bibitem{Hod} O. Hod, V. Barone, and G. E. Scuseria, Phys. Rev. B \textbf{77}%
, 035411 (2008).

\bibitem{EzawaCoulomb} M. Ezawa, Phys. Rev. B \textbf{77}, 155411 (2008).

\bibitem{Wang} W. L. Wang, S. Meng and E. Kaxiras, Nano Letters \textbf{8},
241 (2008). W. L. Wang, \textit{et al.}, Phys. Rev. Lett. \textbf{102},
157201 (2009).

\bibitem{EzawaSpin} M. Ezawa, Eur. Phys. J. B \textbf{67}, 543 (2009)

\bibitem{SchriefferWolff} J. R. Schrieffer and P. A. Wolff, Phys. Rev. 
\textbf{149,} 491 (1966).

\bibitem{Abramowitz} M. Abramowitz and I. A. Stegun, Handbook of
Mathematical Functions with Formulas, Graphs, and Mathematical Tables, New
York: Dover, 1004-1005, (1972).
\end{thebibliography}
\end{document}